# Scheduler Technologies in Support of High Performance Data Analysis

Albert Reuther, Chansup Byun, William Arcand, David Bestor, Bill Bergeron, Matthew Hubbell,
Michael Jones, Peter Michaleas, Andrew Prout, Antonio Rosa, Jeremy Kepner

Massachusetts Institute of Technology

*Abstract*—Job schedulers are a key component of scalable computing infrastructures. They orchestrate all of the work executed on the computing infrastructure and directly impact the effectiveness of the system. Recently, job workloads have diversified from long-running, synchronously-parallel simulations to include short-duration, independently parallel high performance data analysis (HPDA) jobs. Each of these job types requires different features and scheduler tuning to run efficiently. A number of schedulers have been developed to address both job workload and computing system heterogeneity. High performance computing (HPC) schedulers were designed to schedule large-scale scientific modeling and simulations on supercomputers. Big Data schedulers were designed to schedule data processing and analytic jobs on clusters. This paper compares and contrasts the features of HPC and Big Data schedulers with a focus on accommodating both scientific computing and high performance data analytic workloads. Job latency is critical for the efficient utilization of scalable computing infrastructures, and this paper presents the results of job launch benchmarking of several current schedulers: Slurm, Son of Grid Engine, Mesos, and Yarn. We find that all of these schedulers have low utilization for short-running jobs. Furthermore, employing multilevel scheduling significantly improves the utilization across all schedulers.

I. INTRODUCTION

Large assets are often built to be shared among many people and teams. For instance, telescopes and linear accelerators are used for research and experimentation by many researchers and teams. One individual or team is very unlikely to most effectively and consistently utilize such a large asset over time. The economics come out far ahead when such assets are shared. Similar to telescopes and accelerators, large pools of computing capabilities are also shared among many. Multiple users submit a wide variety of computational jobs to be processed on computational resources that include various (and sometimes heterogeneous) processing cores, network links, memory stores, and storage pools along with checking out software execution licenses to run certain licensed applications. Further, each user's job will fit into certain parallel execution paradigms from independent process jobs to independently (pleasantly) parallel to synchronously parallel jobs, each of which imposes certain execution requirements. Job schedulers are one of the most important components of scalable computing infrastructures; if the scheduler is not effectively managing resources and jobs, the computing capabilities will be underutilized.

This material is based upon work supported by the National Science Foundation under Grant No. DMS-1312831. Any opinions, findings, and conclusions or recommendations expressed in this material are those of the authors and do not necessarily reflect the views of the National Science Foundation.

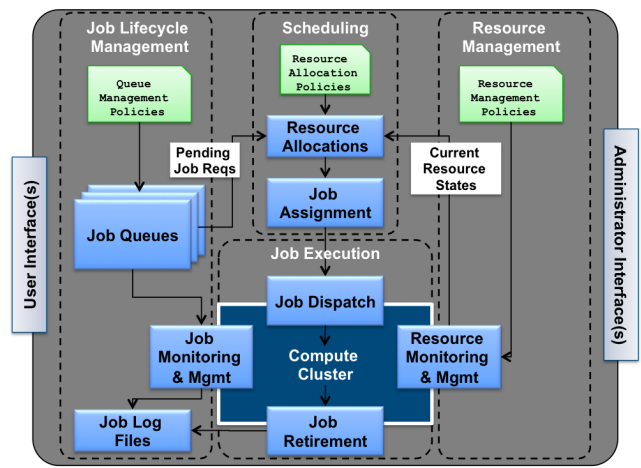

Figure 1: Key components of cluster schedulers including job lifecycle management, resource management, scheduling, and job execution.

Job schedulers go by a variety of names, including schedulers, resource managers, resource management systems (RMS), and distributed resource management systems (D-RMS). These terms are used interchangeably, and we will follow that convention in this paper. The two main terms capture the primary activities of this type of distributed software. At its simplest level, job schedulers are responsible for matching and executing compute jobs from different users on computational resources. The users and their jobs will have different resource requirements and priorities. Similarly, the computational resources have different resource availabilities and capabilities, and they must be managed in such a way that they are best utilized, given the mix of jobs that need to be executed.

In essence, every job scheduler has four key operational tasks: job lifecycle management, resource management, scheduling, and job execution, as shown in Figure 1. The job lifecycle management task receives jobs from users through the user interface and places them in one of the job queues to wait for execution (regardless of whether jobs are scheduled and executed on demand [Reuther 2007] or batch queued). Various resources for the job including memory, licenses, and accelerators (such as GPUs) are requested through the user interface by the user. The job lifecycle management task is also responsible for prioritizing and sorting candidate jobs for execution by using the queue management policies. The scheduling task periodically requests a prioritized list of candidate queued jobs and determines whether resources are available to execute one or more of the jobs. The scheduler receives the state of all the resources from the resource management task, which in turn is receiving resource state and availability information from the compute nodes. The scheduling task allocates resources (usually one or more job slots on compute nodes) and assigns the job to the resource(s) if adequate resources are available to execute each job. The job execution task is responsible for dispatching/launching the job on the resources. Upon the completion of each job, the job execution task manages the closing down of the job and reporting the statistics for the job to the job lifecycle management task, which records it in logs.

This architecture is not much different from those proposed in the early days of parallel computing schedulers; however, the sophistication of resource management, scheduling, resource allocation, queue management, and other features has evolved greatly. This paper surveys these features and presents performance measurements for assessing the suitability of various schedulers for HPDA workloads. The outline of the paper is as follows. Section II overviews the most prominently used schedulers. Section III compares the features of these schedulers and shares the results of job scheduling and execution latency performance benchmarking of the schedulers. Section IV presents related work, and Section V summarizes the paper.

## II. JOB SCHEDULERS

A number of schedulers have been developed over the past thirty-five years to address various supercomputing and parallel data analysis computer architectures, network architectures, and software architectures. One of the first schedulers was the Network Queuing System (NQS) batch scheduler at NASA [Kinsbury 1986]. There are commonalities of intent and features among the job schedulers, and we can categorize current schedulers into several scheduler families.

### A. Scheduler Families

From the time of NQS, high performance computing (HPC) systems have used job schedulers to manage jobs and resources. These schedulers used a batch queue that kept a backlog of jobs to be run, giving the scheduling task many choices from which to choose the optimal next job on the basis of the current and/or near-future resource availability. However, this batch queue required that jobs wait in the queue, often from minutes to days, in order to begin execution. While most HPC centers continue to use batch queues for scheduling jobs, the MIT Lincoln Laboratory team showed that on-demand job launching was possible and desirable for a certain user base [Reuther 2005]. The Big Data community, including Google, Yahoo, Microsoft, and others, found batch queuing job schedulers to be inadequate for their requirements. Big Data jobs tend to be short-duration, independently parallel high performance data analysis jobs and persistent data service jobs, while HPC jobs tend to be long-running, synchronously parallel simulation and modeling jobs.

The MapReduce scheduler, which was among the first Big Data schedulers, is a very simple scheduler, and it was developed because Google MapReduce [Dean 2008] (and consequently Hadoop MapReduce [Dittrich 2012]) jobs required scheduling of jobs to match locally stored files. Subsequently, schedulers like Borg [Verma 2015] and Mesos [Hindman 2011] were developed because of the perception that HPC schedulers could only effectively be used in batch processing modes, and that their main feature was more optimally scheduling jobs that were waiting in batch queues. However, both of these shifts also occurred because of the need for in-language APIs beyond just a command line interface. There was an effort to develop a common in-language API called DRMAA (Distributed Resource Management Application API) for batch schedulers, but adoption of DRMAA was tepid because of nontechnical market factors.

Hence, the main division is between HPC and Big Data schedulers. The HPC scheduler family of schedulers can be further broken into the traditional HPC and the new HPC sub-families. The traditional HPC schedulers include PBS [Henderson 1995], Grid Engine [Slapnicar 2001], HTCondor [Litzkow 1988], OAR [Capit], and LSF [Zhou 1993]. The new HPC schedulers include Cray ALPS [Newhouse 2006] and Slurm [Yoo 2003]. The HPDA schedulers can be further broken into commercial and open-source sub-families. The commercial Big Data schedulers include Google MapReduce [Dean 2008], Google Borg [Verma 2015], and Google Omega [Schwartzkopf 2013], and components of Borg and Omega are available as the open-source Kubernetes project [Burns 2016]. The open-source Big Data schedulers include Apache Hadoop MapReduce [Dittrich 2012], Apache YARN [Vavilapalli 2013], and Apache Mesos [Hindman 2011]. In the next subsection, we will compare representatives from each of the sub-families, namely the following:

- *Grid Engine* is a full-featured, very flexible scheduler matured by Sun Microsystems and Oracle and currently offered commercially by Univa. There are also several open-source versions, including Son of Grid Engine.

- *LSF* is a full-featured and high performing scheduler that is very intuitive to configure and use. OpenLAVA [http://www.openlava.org/] is an open-source derivative of LSF that has reasonable feature parity.

- *Slurm* is an extremely scalable, full-featured scheduler with a modern multithreaded core scheduler and a very high performing plug-in module architecture.

- *Apache YARN* is an in-memory map-reduce scheduler to enable scaling map-reduce style jobs past several thousand servers.

- *Apache Mesos* is a two-level scheduler that enables the partitioning of a pool of compute nodes among many scheduling domains. Each scheduling domain has a pluggable scheduler called a Mesos framework (think queues and node pools), which allocates the resources within its domain resources.

- *Kubernetes* is an open-source distributed Docker container management system that includes a scheduler. It is based on Google Borg and Omega.

### B. Scheduler Features

Depending on the nature of the job policy and primary intent of the large-scale computing system, certain resource manager and scheduler features are key. There are many features that we can analyze among these schedulers; this paper will only compare and contrast the ones that are particularly important to both HPC and HPDA jobs. In particular, we will compare and contrast the following features:

- *Cost and licensing* conveys whether the scheduler is open source or has a license for which one must pay. While it is nice to have contracted support, it can be very expensive for large clusters.
- *Language support* captures the programming languages in which executed applications can be written.
- *Parallel and array jobs* indicates whether the job scheduler accommodates synchronous dependent parallel and/or asynchronously independent parallel (array) job.
- *Job dependencies* means that jobs can be submitted so that they do not start execution until another specified job completes execution.
- *Resource allocation policy* designates whether policy can be described on how resources are allocated to jobs.
- *Scalability and throughput* conveys how many job slots can be supported and managed simultaneously. (Job slots have some reasonable correspondence to the number of cores in the system, though the relationship is not always one to one.)
- *Job restarting* means that the scheduler can automatically restart a job if it has erroneously terminated execution.
- *Job migration* means that the scheduler is able to migrate a job from one compute node to another as it is executing. Usually this feature requires scheduler-based checkpointing to capture the current state of the job so that the current state can be restarted from where it had left off on another node.

*C. Feature Comparison*

Table 1: Key features comparison among job schedulers.

| Feature | LSF | Grid Engine | Slurm | YARN | Mesos | Kubernetes |
|---|---|---|---|---|---|---|
| Type | HPC | HPC | HPC | Big Data | Big Data | Big Data |
| Cost/Licensing | $$$ | $$$, Open source | Open source | Open source | Open source | Open source |
| Language support | All | All | All | Java, Python | All | All |
| Parallel and Array Jobs | Both | Both | Both | Array | Array | Array |
| Job Dependencies | Yes | Yes | Yes | Yes | Some | No |
| Resource Allocation Policies | Yes | Yes | Yes | Yes | Yes | Yes |
| Scalability and Throughput | 10K+ | 10K+ | 100K+ | 10K+ | 100K+ | 100K+ |
| Job Restarting | Yes | Yes | Yes | Yes | Yes | Yes |
| Job Migration | Yes | Yes | Yes | No | User | User |

The comparison of the job schedulers is summarized in Table 1. First, all of these schedulers are actively being developed and are supported on all of the major server-based Linux distributions. Most of the job scheduler offerings are open-source or have an open-source version. The exception is LSF, but LSF does have an open source project called OpenLAVA that has some parity with LSF, though the two schedulers do not share a common code base. Broad language support is almost universal except for YARN. YARN does support other languages to some extent, but Java and Python are the most strongly supported.

The HPC schedulers all support both parallel and array jobs, while the Big Data schedulers support array jobs. Mesos can support parallel jobs, too, when using the MPI framework. Also Mesos supports some job dependency scheduling depending on the framework(s) being used. Kubernetes does not support dependencies.

All of the schedulers support resource allocation policies, and they all support the concept of static and dynamic resources. Static resources are either busy or not (i.e., a job slot), while dynamic resources can be partially used by jobs (i.e., the memory of a compute node). Further, all of the schedulers have some mechanism for administrators to define and manage resources that they specify in configuration scripts.

Each of the schedulers is able to handle anywhere from 10K to over 100K simultaneously executing job slots. This number is primarily dependent on how the job and resource data structures and management algorithms are designed and implemented. Finally, all of the schedulers support job restarting, but job migration is left to the programmers/users in Mesos and Kubernetes.

There is one more key feature essential for HPDA workloads that has not been addressed in this Section: scheduler latency, which is addressed in the next section.

III. PERFORMANCE COMPARISON

If the system is primarily intended for batch processing large, parallel jobs, it is essential that the scheduler is able to schedule these large parallel jobs while also adeptly backfilling shorter, smaller jobs, such as job arrays, in order to keep resource utilization high. That is, HPC schedulers need sophisticated job scheduling algorithms. However, if the system is used mainly to process on-demand interactive jobs that are independently parallel, it is essential that the resource manager quickly updates (with timely periodicity) the status of all of the nodes and that the job scheduler is quick to assign new on-demand jobs to resources. This responsiveness is measured by scheduler latency, which includes the latency of submission, queue management, resource identification, resource selection, resource allocation, job dispatch, and job termination.

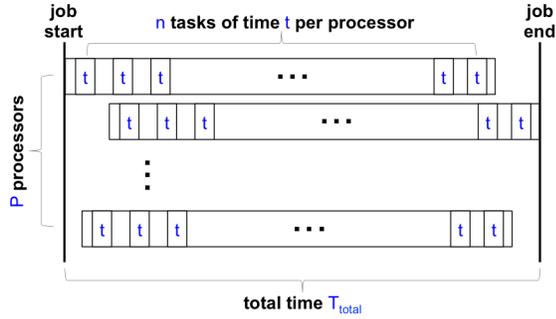

Figure 2: Conceptual scheduler latency experiment test plan.

The experiment test plan is depicted in Figure 2. We schedule $n$ tasks such that each executes in $t$ time on $P$ processor cores. The isolated job execution time is then

$$T_{job} = t \cdot n.$$

We measure the total time to execute the $n$ x $P$ tasks as $T_{total}$. We can then calculate the scheduler utilization as

$$U = T_{job} / T_{total}.$$

This test plan allows us to determine the scaling behavior across schedulers for a large number of short-duration jobs by producing curves of utilization $U$ versus task time.

*A. Benchmarking Environment*

To minimize uncertainty, performance measurements were performed on our MIT SuperCloud cluster [Reuther 2013]. This cluster could be fully isolated from extraneous processes in order to deliver consistent results. Forty-five nodes were used: one to serve as the scheduler node and forty-four as compute nodes (1408 cores total). All of the nodes were connected via a 10 GigE interface to a Lustre parallel storage array.

To compare the launch latency for the schedulers, we chose four representative schedulers from across the scheduler landscape: Slurm, Son of Grid Engine, Mesos, and Hadoop-YARN. Each of the four schedulers was installed on the scheduler node and compute nodes, and the daemons of only the scheduler under test were running, i.e., only one scheduler was running at a time. Each of the schedulers had many tuning parameters, and all four of them were configured and tuned to achieve optimal performance on short tasks while minimizing the impact of these optimizations on other tasks. The version designators and scheduler configuration details are captured in Figure 3.

- **Slurm 15.08.6**
  - `ProctrackType=proctrack/cgroup`
  - `SchedulerType=sched/builtin`
  - `SelectType=select/cons_res`
  - `SelectTypeParameters=CR_Core_Memory`
  - `PriorityType=priority/basic`
  - `DefMemPerCPU=2048`
- **Son of GridEngine 8.1.8**
  - Enabled high-throughput configuration
- **Mesos 0.25.0**
  - Single-node master
  - Single ZooKeeper running on master
- **Hadoop-Yarn 2.7.1**
  - Single-node master running NameNode and ResourceManager (YARN) daemons
  - Compute nodes running DataNode and NodeManager daemons

Figure 3: Scheduler configurations for achieving optimal performance on short tasks.

*B. Latency and Utilization Measurements*

The jobs that were launched on the 44 compute nodes (1408 cores total) through the scheduler were all sleep jobs of 1, 5, 30, or 60 seconds. The total number of tasks $N$ and the number of tasks per processor $n$ were chosen so that the total processor time was always the same: 93.7 hours (337,920 seconds). The four parameter sets that were used for comparison are shown in Table 2. For each parameter set and each scheduler, three trials were executed, and the results are the average of the three trials. With all four schedulers, the jobs were submitted as job arrays.

Each of the parameter sets has only one duration length of jobs. More heterogeneous mixes of job durations can be composed of combinations of these parameter sets (and other in-between duration values); hence, the trends that we learn from these parameter sets can easily be used to deduce the scheduler latency performance of more heterogeneous mixes of job durations.

Table 2: Scheduler latency experimental trials.

| Configuration | Parameter Set 1 | Parameter Set 2 | Parameter Set 3 | Parameter Set 4 |
|---|---|---|---|---|
| Processors P | 1408 | 1408 | 1408 | 1408 |
| Job time per processor $T_{job}$ | 240 secs | 240 secs | 240 secs | 240 secs |
| Task time t | 1 sec | 5 secs | 30 secs | 60 secs |
| Tasks per processor n | 240 | 48 | 8 | 4 |
| Total tasks N | 337920 | 67584 | 11264 | 5632 |
| Total processor time | 93.7 hours | 93.7 hours | 93.7 hours | 93.7 hours |
| Number of trials | | | | |
| Slurm | 3 | 3 | 3 | 3 |
| GE | 3 | 3 | 3 | 3 |
| Mesos | 3 | 3 | 3 | 3 |
| Hadoop-Yarn | | 3 | 3 | 3 |

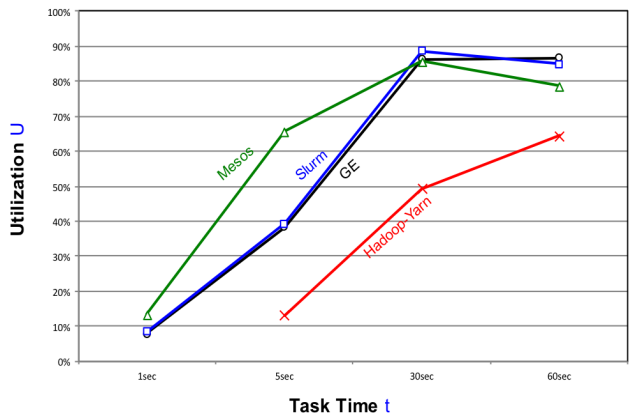

**Figure 4: Utilization comparison of scheduler latencies with varying task times.**

Figure 4 shows the results of launch latency as captured in the utilization defined at the beginning of this section. Since we have a fixed total amount of work, any additional time that it takes the cluster to execute the total set of jobs decreases the utilization of the cluster resources. In other words, if the scheduler consumes a lot of time in one or more of the tasks involved in managing job execution, which are submission, queue management, resource identification, resource selection, resource allocation, job dispatch, and job termination. As we can see in Figure 4, all of the schedulers do well with 60-second tasks; they all launch jobs reasonably fast and fill the cluster quickly to keep it busy. Slurm, Grid Engine, and Mesos perform similarly with 1-, 5-, and 30-second tasks. However, Hadoop-YARN is least efficient among these schedulers – it was so inefficient that the 1-second task trials took too long to run. But even for Slurm, Grid Engine, and Mesos, the lower utilization rates for 1- and 5-second jobs beg for a solution to realize better utilization.

*C. Multilevel Scheduling*

The key to increasing the utilization for 1- and 5-second duration jobs is in decreasing the job launch latency or not launching as many jobs overall while still getting all of the work done. With pleasantly parallel jobs, which most HPDA jobs are, we can modify our analysis code slightly to be able to process multiple datasets or files with a single job launch. We can then use a tool like LLMapReduce [Byun 2016] to efficiently launch the jobs onto the cluster. This technique is referred to as multilevel scheduling.

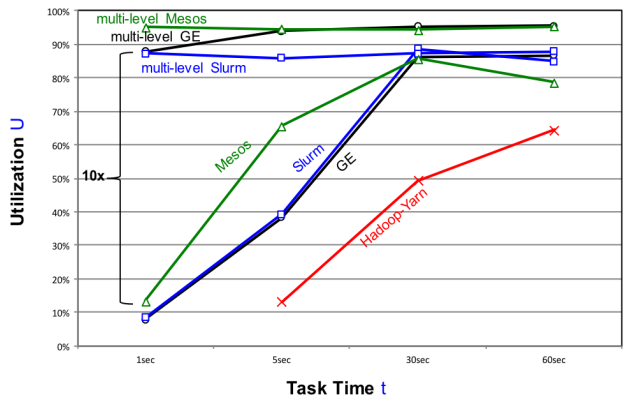

**Figure 5: Utilization comparison of scheduler latencies with varying task times including multilevel scheduling results.**

**Table 3: Scheduler latency experimental trials including multilevel scheduling.**

| Configuration | Parameter Set 1 | Parameter Set 2 | Parameter Set 3 | Parameter Set 4 |
|---|---|---|---|---|
| Processors P | 1408 | 1408 | 1408 | 1408 |
| Job time per processor $T_{job}$ | 240 secs | 240 secs | 240 secs | 240 secs |
| Task time t | 1 sec | 5 secs | 30 secs | 60 secs |
| Tasks per processor n | 240 | 48 | 8 | 4 |
| Total tasks N | 337920 | 67584 | 11264 | 5632 |
| Total processor time | 93.7 hours | 93.7 hours | 93.7 hours | 93.7 hours |
| Number of trials | | | | |
| Multilevel Slurm | 3 | 3 | 3 | 3 |
| Multilevel GE | 3 | 3 | 3 | 3 |
| Multilevel Mesos | 3 | 3 | 3 | 3 |

Table 3 shows the four parameter sets that were used with the LLMapReduce tool to test the utilization of multilevel scheduling on Slurm, Grid Engine, and Mesos. (They are identical to the parameter sets used in Subsection B.) Figure 5 shows the same results as Figure 4 and includes the results of the use of multilevel scheduling. Slurm, Grid Engine, and Mesos each has very high utilization across all task durations with this technique. The figure shows that multilevel scheduling brings the utilization rates for all three schedulers around 90%, which is on par with the 30- and 60-second jobs.

IV. RELATED WORK

In the world of schedulers, a great deal more research has gone into optimizing the placement of jobs onto the resources with a great deal of variety in both the homogeneity and heterogeneity of the jobs and the resources. Many of these studies have put much work into crafting a representative mix of heterogeneous resources (i.e., memory, GPUs, cores, compute power) that best represent the systems they are modeling and a representative mix of job workloads that their systems execute. However, these studies primarily stress the ability of the schedulers to match jobs to resources and only test the launching efficiency of the schedulers. There are several papers, though, that have addressed the efficiency of launching jobs specifically.

Two papers compare job launch time of a monolithic (single) scheduler to that of distributed/partitioned job scheduling. [Brelsford 2012] explores partitioned parallel job scheduling by modifying the IBM LoadLeveler (a modification of HTCondor) scheduler, while [Zhou 2013] explores distributed resource allocation techniques by modifying Slurm. Rather than measuring utilization, both papers measure the throughput of how many jobs they can launch through the scheduler per second. As one might expect, their partitioned parallel and distributed scheduling techniques yielded greater job throughput. A 2014 white paper by the Edison Group [Edison 2014] also uses job throughput to compare several modern HPC schedulers. Finally, we measured launch times of virtual machine jobs onto clusters using a variety of job slot packing and oversubscription techniques [Reuther 2012]. However, we did not make a distinction between the latency of the scheduler and the latency of the virtual machine launching because we were reasonably confident that most of the latency was due to virtual machine launching.

## V. SUMMARY

Schedulers are a key component of a scalable computing infrastructure because they orchestrate all of the work and directly impact system effectiveness. Schedulers have many features, and certain features can enhance the execution of certain workloads. In this paper, we have compared a number of scalable compute cluster schedulers and developed the concept of scheduler families based on their lineage and features. We then compared and contrasted a number of key features and their impact on high performance data analytic workloads. We found that, in general, there was a great deal of support across all of the representative schedulers that we examined.

We then focused on scheduler latency and conducted a series of benchmarks on a further subset of schedulers: Slurm, Grid Engine, Hadoop-YARN, and Mesos. We found that Hadoop-YARN performed worse than the other three schedulers, and that the other three performed similarly. We then used the multilevel scheduling technique with the LLMapReduce tool to get greater cluster utilization from Slurm, Grid Engine, and Mesos.

## ACKNOWLEDGMENTS

The authors wish to acknowledge the following individuals for their contributions: Alan Edelman, Justin Brukardt, Steve Pritchard, Chris Clarke, Sterling Foster, Paul Burkhardt, Victor Roytburd, Dave Martinez, Vijay Gadepally, Anna Klein, Lauren Milechin, Julie Mullen, Sid Samsi, and Chuck Yee.